\documentclass[aps,prb,reprint]{revtex4-1}
\usepackage{epsf}
\usepackage{amsfonts}
\usepackage[fleqn]{amsmath}
\usepackage{amssymb,yfonts,mathrsfs,bbm}
\usepackage{graphicx}
\usepackage{multirow}
\usepackage{longtable}
\usepackage{amsmath}
\usepackage[normal]{caption2}
\usepackage{geometry}
\usepackage{color}
\definecolor{purple}{rgb}{0.7,0.0,0.7}
\definecolor{orange}{rgb}{1,0.65,0.0}
\definecolor{dgreen}{rgb}{0.3, 0.6, 0.1}

\begin{document}
\title{Breakdown of Raman Selection Rules By Fr\"{o}hlich Interaction in Few-Layer WS$_2$}

\author{Qing-Hai Tan$^{1}$}
\author{Kai-Xuan Xu$^{1}$}
\author{Yu-Jia Sun$^{1}$}
\author{Xue-Lu Liu$^{1}$}
\author{Yuan-Fei Gao$^{1,2}$}
\author{Shu-Liang Ren$^{1}$}
\author{Ping-Heng Tan$^{1,2}$}
\author{Jun Zhang$^{1,2*}$}
\affiliation{$^{1}$State Key Laboratory of Superlattices and Microstructures, Institute of Semiconductors, Center of Materials Science and Optoelectronics Engineering, CAS Center of Excellence in Topological Quantum Computation, University of Chinese Academy of Sciences, Beijing 100083, China
\\$^{2}$ Beijing Academy of Quantum Information Science, Beijing 100193, China
\\*Correspondence and requests for materials should be addressed to J. Z. (zhangjwill@semi.ac.cn)}

\begin{abstract}
{The Raman selection rules arise from the crystal symmetry and then determine the Raman activity and polarization of scattered phonon modes. However, these selection rules can be broken in resonant process due to the strong electron-phonon coupling effect. Here we reported the observation of breakdown of Raman selection rules in few-layer WS$_2$ by using resonant Raman scattering with dark A exciton. In this case, not only the infrared active modes and backscattering forbidden modes are observed, but the intensities of all observed phonon modes become strongest under paralleled-polarization and independent on the Raman tensors of phonons. We attributed this phenomenon to the interaction between dark A exciton and the scatted phonon, the so-called intraband Fr\"{o}hlich interaction, where the Raman scattering possibility is totally determined by the symmetry of exciton rather than the phonons due to strong electron-phonon coupling. Our results not only can be used to easily detect the optical forbidden excitonic and phononic states but also provide a possible way to manipulate optical transitions between electronic levels. }
\end{abstract}

\maketitle
\section{INTRODUCTION}
Besides the energy and momentum conservation, the Raman scattering processes must follow the so-called Raman selection rules that governed by parities of phonon wavefunctions and hence the symmetry of materials\cite{Ramaneffect}. These selection rules are very useful for investigating the Raman activity and polarization property of scattered phonons. Based on Raman selection rules, the infrared (IR) active and Raman active phonons are complementary in centrosymmetric crystals; in some crystals, there are some silent phonon modes which are neither IR nor Raman active, or forbidden in certain scattering geometry. However, these selection rules will be broken under special conditions, such as the electronic field gradient effects of metal plasmonic structure\cite{PRL-EFGR-2000,Mai2013Selection} and resonant Raman scattering (RRS) due to strong electron-phonon coupling (EPC)\cite{cu2o-prl-1973,Martin1983,tqh-2D}. The EPC plays an important role in thermodynamics, electronic transport and optical properties of solids. Manipulation of EPC will give rise big opportunity to discover new physics phenomena and design novel electronic and optoelectronic devices. Two-dimensional materials (2DMs) and related 2D van der Waals heterostructures (vdWHs) assembled by different 2DMs provide a perfect platform to engineer the EPC, thus to design novel optoelectronic devices\cite{cuixiaodong-review-2015,C8CS00332G,Zhang-CSR-2015}. In such 2D system, the EPCs can be dramatically modified and generate many novel physical effects, such as unconventional superconductivity in twisted bilayer graphene\cite{magicgraphene-1,magicgraphene-2,magic-prl}, quantum entanglement between chiral phonon and single photon in WSe$_2$\cite{Ajit-NP}, cross-dimensional EPC in vdWHs\cite{MLL-NC-2019}, dark-exciton resonance in WS$_2$\cite{tqh-2D}, and improving degree of valley polarization by suppress the electron-phonon coupling\cite{Frohlich-NC-2019}. It is particularly significant to explore Raman selection rules of 2DMs in strong electron-phonon coupling regime, but still elusive so far.

Here we investigated the breakdown of Raman selection rules in few-layer WS$_2$ by polarization dependent resonant Raman spectroscopy. Multiple laser lines closed to the energies of A, B, and C excitons (electron-hole pairs) of few-layer WS$_2$ were used to evaluate the resonant behavior of phonons. When the excitation energy is slightly below bright A exciton state and close to dark A exciton state, a series of IR-active modes and normally forbidden modes in backscattering configuration were clearly observed. In particular, the polarization behaviors of all modes become parallel polarization and independent on the Raman tensor of modes. Experimental and theoretical analysis show that such a breakdown of Raman selection rules originates from intraband Fr\"{o}hlich interaction between the dark A excitons and scattered phonons.

\begin{figure}[htb]
\centerline{\includegraphics[width=88mm,clip]{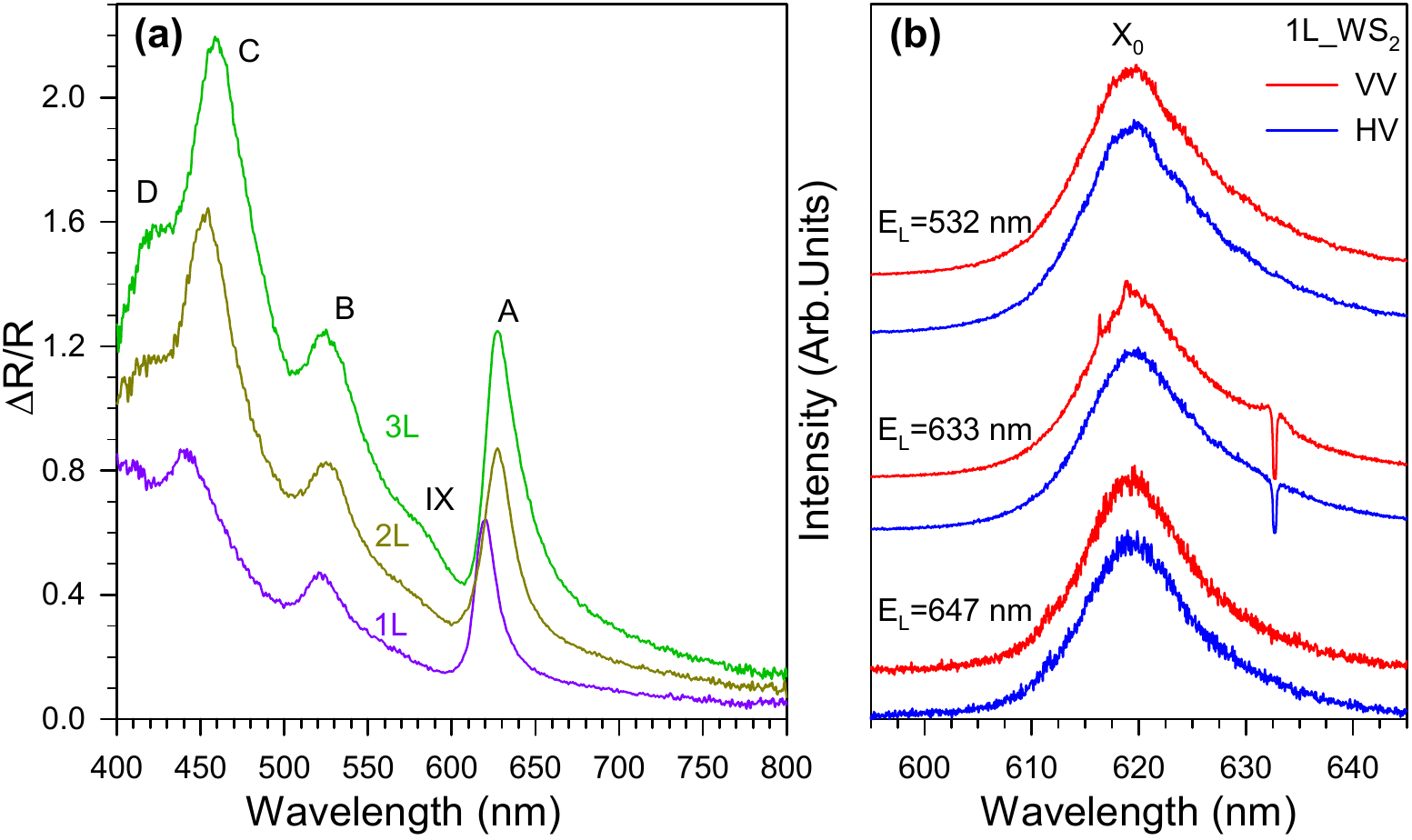}}
\caption{(a) Reflectance contrast spectra $\Delta{R/R}$ of 1L, 2L, and 3L WS$_2$ crystals on SiO$_2/$Si substrate. (b) The polarized PL spectra of 1L WS$_2$ with excitation energies close to B, A exciton and slightly below bright A exciton, respectively.} \label{Fig1}
\end{figure}

\section{EXPERIMENTAL SECTION}
The samples were prepared from bulk WS$_2$ crystals onto a 90 nm SiO$_2/$Si substrate by using the micromechanical exfoliation technique. Raman measurements were undertaken in backscattering geometry with a Jobin-Yvon HR800 system equipped with a liquid-nitrogen-cooled charge-coupled detector. The spectra were collected with a $\times$100 objective lens (NA=0.9) and an 1800 lines mm$^{-1}$ grating at room temperature. The excitation laser (E$_L$) lines of 457 nm and 488 nm are from an Ar$^{+}$ laser; the laser lines of 612 nm and 633 nm are from a He-Ne laser; the laser lines of 532 nm, 647 nm and 676 nm are from a Kr$^{+}$ laser. The ultralow-frequency Raman spectra were obtained down to $\pm$5 cm$^{-1}$ by combining three volume Bragg grating filters into the Raman system to efficiently suppress the Rayleigh signal. In order to avoid the laser heating effect to the samples, the laser power was kept below a maximum of 0.2 mW. The reflectance contrast $\Delta{R/R}$ were undertaken with a $\times$100 objective lens (NA=0.9) and an 100 lines mm$^{-1}$ grating with white light at room temperature.

\section{RESULTS}
Figure 1(a) shows the reflectance contrast spectra $\Delta{R/R}$ of 1-3L WS$_2$ at room temperature\cite{tqh-2D}. Two features at around 627 nm (1.98 eV) and 528 nm (2.35 eV) are denoted by A and B exciton, which originates from direct transitions between the spin-orbit split valence band and the conduction at K (or K$'$) point of the Brillouin zone\cite{Mark-F-prl-2010}, respectively, the C peak at around 460 nm (2.7 eV) is from transition between the highest valence band and the lowest conduction bands around the $\Gamma$ point of the Brillouin zone\cite{Qiu-PRL-2013}. The peak at around 576 nm (2.15 eV) is denoted by IX exciton may correspond to the interlayer exciton of WS$_2$ or the excitonic Rydberg state of A exciton\cite{IX-PRB}. The origin of D peak at around 413 nm (3.0 eV) is not clear so far. Owing to the direct band gap of monolayer (1L) WS$_2$, the energies of A, B and C excitons in 1L are slightly larger than those in bilayer (2L) and trilayer (3L) WS$_2$, as shown in Fig. 1(a). Moreover, in contrast to the A and B excitons, C exciton is much less confined to a single layer WS$_2$\cite{Carvalho-prl-2015,PhysRevB-resonant-2015}. Figure 1(b) shows the polarized photoluminescence (PL) spectra of bright A exciton (X$_0$) in 1L WS$_2$ with three different laser lines, where one is downconversion PL excited by 532 nm (2.34 eV), another two are upconversion PL excited by 633 nm (1.96 eV) and 647 nm (1.92 eV), respectively. Obviously, the intensities of X$_0$ peaks are almost the same under parallel (VV) and cross (HV) polarization configurations, implying the bright A exciton is isotropic. It also indicates that its emission and absorption are depolarized for linearly polarized laser. Beside the bright excitons, the conduction band also has a small splitting by tens of meV, which causes the dark A and B exciton transitions\cite{Pola-PRB-2015,Splitting-PRB-2016}. Generally, these dark states lying below the bright states for WS$_2$ (X=S, Se), while they are reversed for MoX$_2$\cite{Pola-PRB-2015,Splitting-PRB-2016}. Actually, the dark A exciton of WX$_2$ has been directly observed by different experiments methods\cite{dark-2DM-2017,philipkim-nn-2017,PhysRevLett-dark,dark-purcell-nn}.

\begin{figure}[htb]
\centerline{\includegraphics[width=90mm,clip]{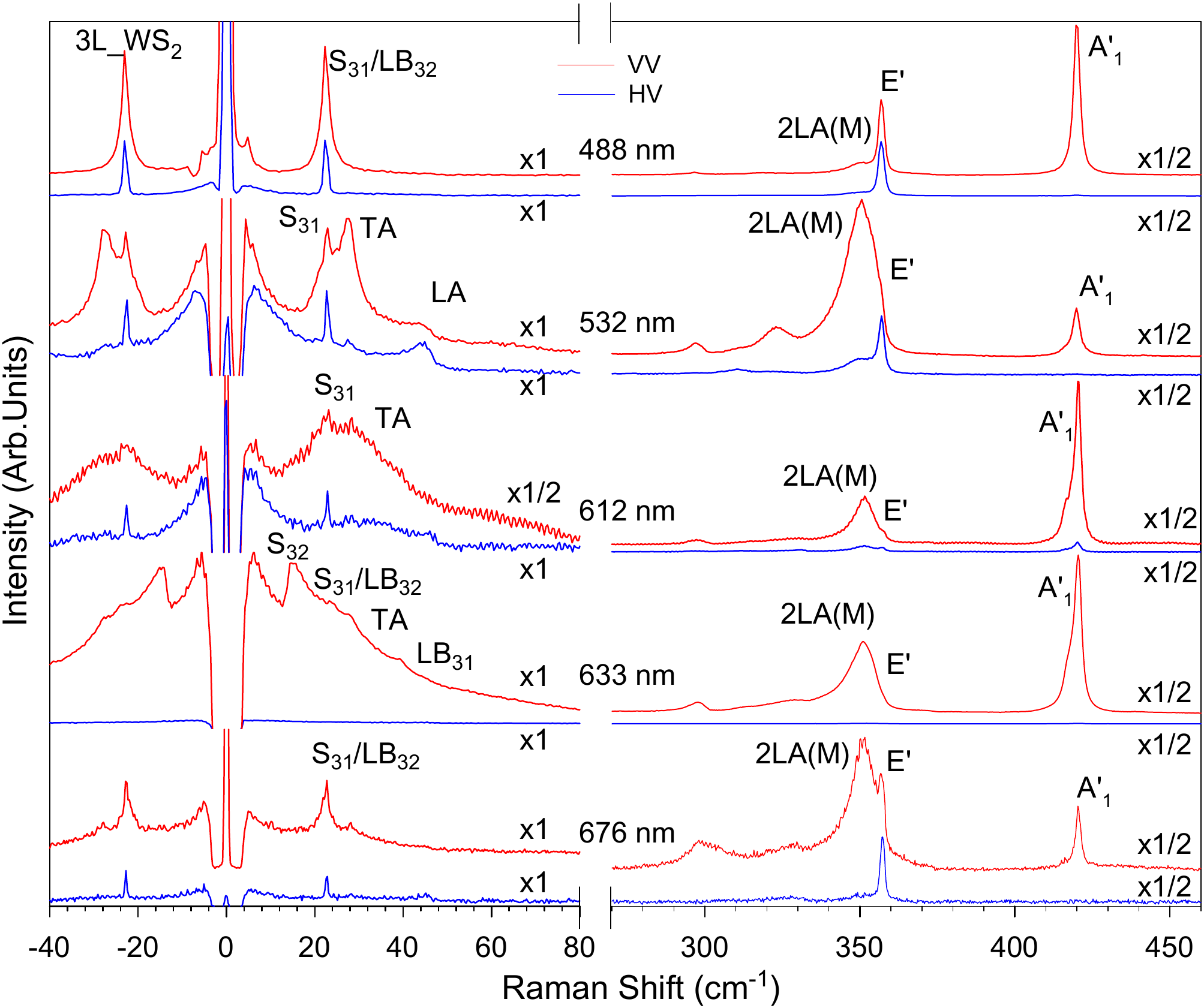}}
\caption{The polarized Raman spectra of 3L WS$_2$ with two excitation wavelengths close to C, B exciton and other three excitation wavelengths slighter above, close to and below bright A exciton, respectively.} \label{Fig2}
\end{figure}

To study the polarization behaviors of resonant Raman spectra, we performed the polarized Raman measurements at room temperature with different excitation wavelengths. The intensity of Raman mode can be described as: $I \propto |\mathbf{e}_s\cdot{\mathbf{\mathcal{R}}}\cdot{\mathbf{e}_i}|$, where $\mathcal{\mathbf{\mathcal{R}}}$ is the Raman tensor of the vibration mode, $\mathbf{e}_i$ and ${\mathbf{e}}_s$ are the polarization vectors of the incident and scattered photons, respectively. Considering the backscattering configurations, ${\mathbf{e}}_i ({\mathbf{e}}_s)$ can only be along x or y direction, and the corresponding matrix representations are $(1, 0, 0)$and $(0, 1, 0)$, respectively. The polarization of WS$_2$ Raman modes is determined by their Raman tensors. The detailed descriptions of Raman tensors of all modes in WS$_2$ are presented in Supplementary Information. Figure 2 shows the Raman spectra of 3L WS$_2$ under parallel (VV) and cross (HV) polarization configurations, where the excitation energies are close to C, B, and A exciton, respectively. Obviously, under resonance with B or C exciton energies, both the in-plane interlayer shear (S) mode and intralayer E$'$ mode survived under VV and HV polarization configurations, whereas out-of-plane interlayer layer breathing (LB) mode and intralayer A$'_1$ mode only survived under VV configurations, which are consistent with the Raman tensor of these modes. In addition, the LA(M) (177 cm$^{-1}$) and LA(M)+TA(M) (311 cm$^{-1}$)\cite{ShiW-2D-2016} modes are also been observed under both VV and HV polarization configurations, as shown in Figure S1 to Figure S3. It indicates these two modes are not totally polarized under resonance with B or C exciton energies. Remarkably, when the excitation energies are close to bright A exciton energy (around 1.98 eV), some abnormal phenomena were observed. Specifically, when the excitation energy is slightly higher than or far below than bright A exciton energy, for example, 612 nm (2.03 eV) and 676 nm (1.83 eV), these modes still comply with the usual Raman selection rules, similar to the case where the excitation energy is resonant with C and B excitons. However, when the excitation energy is slightly below the bright A exciton energy, such as the excitation using 633 nm (1.96 eV) and 647 nm (1.92 eV) (also see in Figure S4 and S5 of Supplementary Information), besides the Raman active modes, the IR-active (e.g. LB$_{31}$) and normally backscattering forbidden modes (e.g. both IR and Raman active S$_{32}$) appear with very strong intensity. The in-plane shear modes (S$_{32}$ and S$_{31}$) shows Fano line-shape. These phenomena can be well explained based on the parity selection rules due to resonant Raman process involved with dark A exciton\cite{tqh-2D}. Remarkably, all of the Raman modes show a strong intensity under parallel polarization configurations, but disappear under cross polarization configurations, including LA(M), and in-plane vibration S$_{31}$ and E$'$ modes. For other layers, similar results were also obtained under this resonance condition, as shown in Figure S4 and Figure S5. These abnormal polarization behaviors indicate that Raman selection rules were broken under special resonance conditions.

\begin{figure*}[htb]
\centerline{\includegraphics[width=140mm,clip]{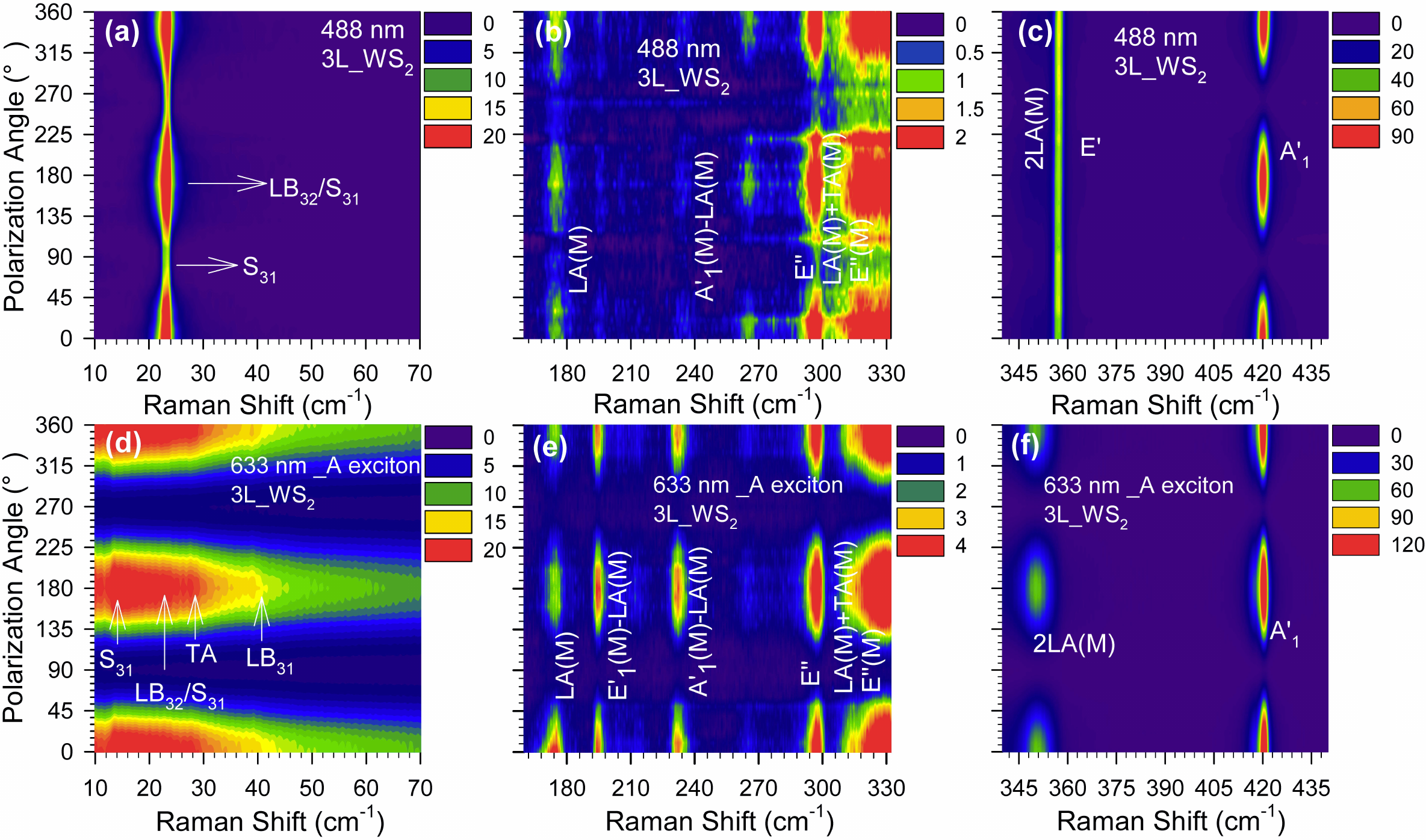}}
\caption{(a-c) Polarization-resolved Raman intensities of Raman modes in ultralow frequency to high frequency region (including shear mode (S$_{31}$), layer breathing mode (LB$_{31}$), and E$'$ and A$'_{1}$ modes) in 3L WS$_2$ under 488 nm, respectively. (d-e) Polarization-resolved Raman intensities of Raman modes in ultralow frequency to high frequency region in 3L WS$_2$ under 633 nm. To clearly see the evolution of intensities with polarization angle, here we present the data into three separate figures.} \label{Fig3}
\end{figure*}

To further study and confirm this abnormal phonon polarization behavior, we conducted polarization-resolved Raman spectra of 3L WS$_2$ excited by 488 nm and 633 nm, respectively. We changed the polarization of incident laser by a half-wave plate and fixed the polarizer at the vertical polarization configuration for scattering signals. Figure 3(a-c) show the polarized Raman spectra of 3L WS$_2$ excited by 488 nm laser. Obviously, the intensities of shear mode and E$'$ mode are independent on the polarization of incident light, while the intensity of A$'_1$ mode is angle-dependent on the polarization of incident light. The evolution of intensities of A$'_1$ modes with polarization angle can be described by $I \propto \cos^2(\beta)$, where $\beta$ is the polarization angle of excitation laser away from y axis. Figure 3(d-e) show the polarized Raman spectra of 3L WS$_2$ excited by 633 nm, closing to the energy of dark A exciton. Interestingly, in this case, all modes show the same angle-dependent behavior, including in-plane vibration shear modes, TA and LA(M) modes, and the 2LA(M) and A$'_1$ modes. These results indicate that under 633 nm resonant excitation condition, polarization behaviors of all Raman modes in WS$_2$ do not follow the symmetry of Raman tensors of vibration modes.

\section{DISCUSSION}

As shown above, the forbidden Raman modes and abnormal polarization behavior only were observed under resonance with dark A exciton rather than dark B exciton. This is because that the dark A states are slightly mixed states with bright A exciton. Consequently, the oscillator strength of dark A exciton is quite small but not equal to zero, while the oscillator strength of dark B is strictly equal to zero\cite{Splitting-PRB-2016,zhangXX-nn-2016}.

To understand the abnormal polarization behavior of scattered phonons in WS$_2$ under this resonant condition, we need to analyze it from the first-order Raman scattering cross section\cite{PRL-break-1971,Martin1983,PRB-P-RRS-1971}:
\begin{equation}
\begin{split}
&\sigma = \sigma_0\frac{\omega_s}{\omega_i}|\sum_{\alpha\beta}{{\mathbf{e}^{\alpha}_{s}}}{{\mathcal{R}}}^{\alpha\beta}{{\mathbf{e}}^{\beta}_i}|^2\\
&{\mathcal{R}}^{\alpha\beta} = \sum_{ij}\frac{{P}^{\alpha}_{0j}{M}_{ij}{P}^{\beta}_{i0}}{(E_{j}-\omega_i+\omega_0)(E_i-\omega_s)}\\
\label{Equation 1}
\end{split}
\end{equation}
where $\sigma_0$ is the free-electron Compton cross section $(e^2/mc^2)^2$ and $\alpha$ and $\beta$ are Cartesian indices, $\omega_{s(i)}$ denotes the frequency of scattered (incident) photons, and $E_i$ and $E_j$ denote the energies of excited electronic states, ${P}^{\alpha}_{0i}$ and ${P}^{\beta}_{0j}$ are the momentum matrix element with the ground state $0$, ${M}_{ij}$ is the phonon scattering matrix element, which both the exciton-phonon interaction and exciton-photon interaction are included, $\omega_0$ is the phonon frequency, and ${\mathbf{e}}_i$ and ${\mathbf{e}}_s$ are incident and scattered photon polarization. Because the tensorial character of $\mathbf{\mathcal{R}}$ determines the polarization selection rules, so we need to consider the expression of $\mathbf{\mathcal{R}}$ under resonant condition.

The electron in bright A exciton state will extremely quickly transit to the dark A state by intravalley spin-flip scattering process with a much faster time scale than radiative recombination, resulting in the low PL quantum yield of monolayer MX$_2$\cite{afm-radiative-2016,nanolett-spinflip-2018}. In particular, the dark A exciton has a longer radiative lifetime, i.e., two orders of magnitude longer than the radiative lifetime of the bright exciton\cite{PRB-DARKLT-2017}. It will cause the accumulation of electrons in dark A exciton state. Therefore, exciton intraband scattering is dominated by dark state band rather than by bright state band. The dark exciton-phonon interactions here can be described by using intraband Fr\"{o}hlich interaction\cite{Martin1983,PRB-P-RRS-1971}. The Hamiltonian $\mathbf{H}_{(F,q)}$ of Fr\"{o}hlich interaction can be written as\cite{Yu2010}
\begin{equation}
\mathbf{H}_{(F,q)}=iC_F/{q}[e^{(ip_h{\mathbf{q}}\cdot{{\mathbf{r}}})}- e^{(ip_e{\mathbf{q}}\cdot{{\mathbf{r}}})}](a^{+}_{k+q}a_k)(C^{+}_{-q}+C_q)
\end{equation}
where $C_F=e[\frac{2\pi\hbar\omega_0}{NV}(\varepsilon^{-1}_{\infty}-\varepsilon^{-1}_0)]^{1/2}(4\pi\varepsilon_0)^{-1/2}$ is a constant coupling coefficient, N and V are the number of unit cells per unit volume of the crystal and the volume of the primitive cell, respectively. $\mathbf{q}=\textbf{k}_i-\textbf{k}_s$ is the wave vector of phonon, and $\varepsilon_0$ is the low-frequency optical dielectric constant and $\varepsilon_{\infty}$ is the static dielectric constant. $C^{+}_q$ and $C_q (a^{+}_k$ and $a_k$) correspond to the creation and annihilation operators for phonons (excitons), respectively. The $\mathbf{r}$ is a characteristic length of the excited state. Since the intraband Fr\"{o}hlich interaction here connects the $s$ states of dark A exciton, the magnitude of the phonon scattering can be simply written as\cite{PRL-break-1971}:
\begin{equation}
\begin{split}
&{M}_{ij} = |\langle{1s}|\mathbf{H}_{(F,q)}|1s\rangle|\\
&= (\frac{C_F}{q})[(\frac{1}{1+(p_h {qr}/2)^2})^2-(\frac{1}{1+(p_e{qr}/2)^2})^2]
\end{split}
\end{equation}
where $p_e=\frac{m_e}{m_e+m_h}$, $p_h=\frac{m_h}{m_e+m_h}$. For the small but nonnegligible ${q}$, this magnitude expression can be expanded as
\begin{equation}
{M}_{ij} \simeq C_F{{qr}}\frac{m_e-m_h}{m_e+m_h}
\end{equation}
Since ${M}_{ij}$ is a constant only for ${i=j}$ and is zero for $i \neq j$, the original Raman tensor $\mathbf{\mathcal{R}}$ in equation (1) degenerates into a scalar quantity: ${{\mathcal{R}}\propto{qr}}$. This result means that the Raman selection rules are no longer restricted according to Raman tensors and become isotropic (more details are present in Supplementary Information). In this case, if we changed the polarization of incident photon and fixed the polarization of collected signals, i.e., $\mathbf{e}_i$ and $\mathbf{e}_s$  are $[\sin{\beta}, \cos{\beta}, 0]$ and $[0, 1, 0]$, respectively, then the Raman intensity can be simply written as:
\begin{equation}
\begin{split}
&I \propto ({qr})^2\cos^2{\beta}\\
\end{split}
\end{equation}
It means the intensity of Raman modes is contributes only to diagonal scattering and irrespective of the Raman tensor of phonons. When $\mathbf{e}_i$ $\parallel$ $\mathbf{e}_s$, the Raman intensity reaches the strongest. Obviously, these analyses perfectly explain our polarization dependent experimental results.

\vspace*{5mm}
\section{CONCLUSIONS}
In summary, we studied the breakdown effect of strong EPC on Raman selection rules in few-layer WS$_2$. Under dark A exciton resonance, not only a series of IR-active and backscattering forbidden vibration modes are active, but also polarization behaviors of all phonon modes are independent on phonon symmetries. We explained such breakdown of Raman selection rules as intraband Fr\"{o}hlich interaction of dark A exciton and phonons. In this resonant process, Raman tensor mainly determined by intermediate exciton transition and degenerated into a scalar quantity format, which leads a parallel polarization behavior of all scattered phonon modes. All theoretical results are independent on the crystal symmetry and expected to be observed in other 2DMs and vdWHs. This breakdown the selection rules not only can modulate Raman intensity and polarization, but also increases the optical excitation channels beyond the restriction of selection rules. These characteristic features are expected to provide the significant advantage of controlling the EPC in semiconductors.

\vspace*{5mm}
\noindent {\bf \ Acknowledgements}
\noindent J.Z. and P.T. acknowledge support from National Basic Research Program of China (grant no. 2017YFA0303401, 2016YFA0301200). Beijing Natural Science Foundation (JQ18014), and Strategic Priority Research Program of Chinese Academy of Sciences (Grant No. XDB28000000).
\vspace*{4mm}

\noindent {\bf \ Author Contribution}
\noindent J.Z. and Q.T. conceived the ideas; Q.T., P.T. and J.Z. designed the experiments. Q.T., X.L., S.R. and Y.S prepared the samples. Q.T. performed experiments. Q.T. and Z.J. analyzed the data and wrote the manuscript with inputs from all authors.

\vspace*{4mm}
\noindent {\bf \large Additional information}

\noindent {\bf Supplementary Information} is available in the online version of the paper. Correspondence and requests for materials should be addressed to J. Z.(Email: zhangjwill@semi.ac.cn).

\vspace*{5mm}
\noindent {\bf Competing interests:} The authors declare no competing financial interest.

\vspace*{8mm}

\end{document}